# Ultrafast graphene photodetector

Fengnian Xia[*‡], Thomas Mueller[‡], Yu-ming Lin, Alberto Valdes-Garcia and Phaedon Avouris[*]

*IBM Thomas J. Watson Research Centre, Yorktown Heights, NY 10598*

**Abstract.** The electronic properties of graphene are unique and are attracting increased attention to this novel 2-dimensional system[1-6]. Its photonic properties are not less impressive[7-8]. For example, this single atomic layer absorbs through direct interband transitions a considerable fraction of the light, $\pi\alpha$ (~2.3%) where $\alpha$ is the fine structure constant, over a very broad wavelength range[7]. However, while applications in electronics are vigorously being pursued, photonic applications have not attracted as much attention. Here, we report on ultrafast photocurrent response measurements in graphene (single and few-layers) field-effect-transistors (FETs) up to 40 GHz light intensity modulation frequencies, using a 1.55 μm excitation laser. No photoresponse degradation is observable up to the highest measured frequency, demonstrating the feasibility and unique benefits of using graphene in photonics. Further analysis suggests that the intrinsic bandwidth of such graphene FET based photodetectors may exceed 500 GHz. Most notably, the generation and transport of the photo-carriers in such graphene photodetectors are fundamentally different from those in currently known semiconductor photodetectors, leading to a remarkably high bandwidth, zero source-drain bias (hence zero dark current) operation, and good internal quantum efficiency.

[‡]*These authors contributed equally to this work*
[*]*Email: fxia@us.ibm.com; avouris@us.ibm.com*

Graphene is a single atomic layer 2-dimensional system composed solely of carbon atoms arranged in a hexagonal honeycomb lattice. It has a linear energy dispersion relation, implying a vanishing effective mass, a high Fermi velocity (~1/300 of the speed of light), and a huge electrical mobility approaching 200,000 cm$^2$/Vs for a free sheet for both electrons and holes [1-5]. Moreover, the photonic properties of graphene are equally remarkable: despite being only a single atomic layer thick, it can absorb photons at least from the visible to the infrared range and the strength of the inter-band transition in graphene is one of the largest in all known materials [7-8]. By comparison, ultra-thin (< a few nanometers) III-V quantum wells can be transparent to a broad range of infrared light due to the blue shift of the absorption edge induced by energy quantization [9]. Multiple graphene layers absorb additively and the absorption range of the system can be tuned by changing the Fermi energy using an external gate field [8, 10].

The combination of the exceptional transport and optical properties points to novel photonic devices fundamentally different from the conventional ones. It has been proposed previously to use graphene nano-ribbons possessing a bandgap as phototransistors [11]. Here we explore the use of zero band-gap, large-area single or few-layers of graphene FETs as ultrafast photodetectors. Upon light absorption, the generated electron-hole pairs in graphene would normally recombine on a time scale of tens of pico-seconds, depending on the quality and carrier concentration of the graphene [12-14]. If an external field is applied, the pairs can be separated and a photocurrent is generated. The same happens in the presence of an internal field. Such fields have been shown to be formed near the metal electrode-graphene interfaces [15-17]. In this paper, we demonstrate that these internal fields can be used to produce ultrafast photocurrent response in graphene. Due to the high carrier transport velocity even under a moderate E-field, no direct bias voltage between source and drain (photocurrent generation path) is needed to ensure ultrafast and efficient (6 to 16% internal quantum efficiency within the photo-detection region) photocurrent generation. Previously, such a zero-bias operation was demonstrated in special uni-travelling-carrier photodiodes (UTC-PD) with much higher built-in E-field for a high bandwidth operation [18-19]. However, in most conventional photodiodes, large external bias directly applied to the photo-detecting area and its complete depletion are essential for fast and efficient photo-detection [20]. Graphene photodetectors may find a wide range of photonic applications including high speed optical communications, interconnects, terahertz detection, imaging, remote sensing, surveillance, and spectroscopy [20-24].

Figure 1a shows a scanning electron micrograph of graphene devices. The optical microscope image in the inset shows the same graphene devices at lower magnification. The number of graphene layers is first estimated from the color of the optical images and then confirmed by Raman spectroscopy [25]. The $SiO_2$ film underneath the graphene is 300 nm thick, and the substrate is highly resistive silicon (1 to 10 kΩ•cm), in order to minimize the parasitic capacitance. Two types of high-frequency coplanar waveguide wirings are shown: ground-signal (G-S) and ground-signal-ground (G-S-G) [26]. The results presented in the main text are obtained from a device with 2 to 3 layers of graphene in the G-S wiring scheme. Results from single layer graphene are presented in the Supplementary Information. The fabrication and layout of the devices are discussed in the Methods section.

A high frequency electrical model of the graphene device is shown in Fig. 1b. It can be modeled by the pad capacitance ($C_p$) in parallel with the graphene capacitance ($C_g$) and graphene resistance ($R_g$), as shown by the green symbols from top to bottom in Fig. 1b. The graphene capacitance $C_g$ quantifies the correlation between the total channel charge and the source-drain bias. The total high frequency impedance, $Z_{IN}$, is:

$$Z_{IN} = \frac{R_g}{1+j2\pi f R_g (C_p + C_j)} \quad (1)$$

where f is the frequency of the excitation electric signal. The impedance can be determined from the measured high frequency reflection coefficient, $\Gamma_{IN}$, from which the resistive and capacitive components are inferred [27]. Details regarding the high frequency calibration and measurement are presented in the Methods section. The measured total capacitance $C_p+C_g$ and $R_g$ in the high frequency domain is plotted in Fig. 1c as a function of the gate bias. Also shown in Fig. 1c is the DC resistance measured between the source and drain of the same graphene device. The graphene resistance measured using the reflection approach in the high frequency domain describes the gate dependence well with a maximum deviation of ~15% from the corresponding DC measurements. Separate measurements on naked G-S contact pads reveal a gate bias insensitive $C_p$ value ranging from 20 to 25 fF, indicating that the graphene capacitance, $C_g$, strongly depends on the gate bias and therefore channel Fermi-level [28].

Photocurrent generation experiments are performed at both low (100 Hz or DC) and high (10 MHz to 40 GHz) light intensity modulation frequencies. Figure 2a shows typical DC current measurements as a function of the source-drain bias with and without light illumination at a fixed gate bias. The dark current is represented by the dashed line which extrapolates through the origin. In this work, the photocurrent is always studied at or close to a short-circuit condition. The magnitude of the photocurrent strongly depends on the location of optical illumination and also on the gate bias as reported previously [16]. Hence, photocurrent imaging experiments at a lock-in frequency of 100 Hz are carried out to determine the optimal illumination locations at various gate biases [16]. Usually, the strongest photocurrent is observed when light is incident near the metal-graphene interfaces where the strongest E-fields form in our p-type FETs. However, at a large positive gate bias, the middle of the channel turns n-type while the graphene in a vicinity of the metal remains p-type due to metal-induced doping. Thus, graphene p-n junctions are formed on both sides of the device and the maximum E-field moves towards the center of the channel for up to 0.5 μm [16-17]. The peak DC photo-responsivity [9], $R_{DC}$ defined as $I_{ph}/P_{opt}$, where $I_{ph}$ is the photocurrent and $P_{opt}$ is the total incidence optical power on the graphene, as a function of gate bias, is shown by the red dots in the inset of Fig. 2b. A maximum external DC photo-responsivity of ~0.5 mA/W is achieved at a gate bias of 80 V.

The reported external efficiency of our device is not intrinsic to the graphene photodetector and can be improved considerably. There are two major factors which limit the external efficiency. The first one is the limited absorption of the light within few-layers of graphene in the current device configuration. A suspended bi- or tri-layer graphene in air absorbs about 4.6% and 6.9% of the vertical incidence light, respectively. In this experiment, the graphene is on top of a 300 nm thick oxide on silicon substrate. At 1.55 µm, although the absorption is estimated to be enhanced by about 25%, it is still far from complete absorption. The second is the width of the effective photo-detection area of the graphene device (or high built-in E-field region), which is only about 100 to 250 nm long (determined by the charge transfer between the metal and graphene) [16]. Given that the light spot size is around 2 µm in diameter, the estimated internal quantum efficiency (the ratio of the number of photo-carriers collected in the external circuits to the number of photons actually absorbed in the small graphene photo-detection area) within the high E-field photo-detection region is 6 to 16%. The efficiency can be further enhanced by creating a wider photo-detection region [16] using split gates, a longer light-graphene interaction length through waveguide [21] or cavity integration, or simply by reducing the internal resistance $R_g$. Furthermore, if a bi-layer graphene is integrated on top of a 550 nm by 220 nm silicon optical waveguide, our numerical simulation indicates that the TM (transverse magnetic) light in such an integrated system can be absorbed by 65% after propagating in the waveguide for 20 µm. Recent experiments on graphene p-i-n junctions we also found considerable improvement in external responsivity (efficiency). Detailed discussions about the schemes and results of improving the external responsivity of graphene photodetectors will be presented elsewhere.

Following the low frequency measurement of the photocurrent, we performed high frequency photo-response measurement from 10 MHz to 40 GHz using a commercial light-wave component analyzer, which measures the peak AC photo-responsivity as a function of the light intensity modulation frequency. Detailed information about the measurements can be found in the Methods section. Figure 2b shows the relative AC photoresponse of the graphene device, $S_{21}(f)$, as a function of the light modulation frequency f up to 40 GHz at a gate bias of 80 V. Here, $S_{21} = 20 \times \log_{10}\left[\dfrac{R_{AC}(Amp/Watt)}{1(Amp/Watt)}\right]$, $R_{AC}$ is the AC photo-responsivity defined as $\Delta I_{ph}/\Delta P_{opt}$, where $\Delta I_{ph}$ and $\Delta P_{opt}$ are the variation of the photocurrent and incident optical power, respectively. The curve is intentionally displaced vertically so that the response starts at 0 dB. No photo-response degradation from the graphene device is observed up to our measurement limit of 40 GHz. The ~1 dB response degradation seen in Fig. 2b is not introduced by the graphene device, but rather by the microwave probes as specified by the manufacturer. The absolute AC photo-responsivity of the same graphene device is measured using a different setup with maximum operational frequency up to 26 GHz. This setup has lower noise hence it allows for the accurate measurement of AC photo-responsivity at all gate biases. A number of frequency response curves are measured and plotted in Fig. 2c and its inset, and no measurable photo-response degradation is observed in all gate biases. The amplitude of the high frequency photo-response does show strong gate dependence, which is similar to that in DC measurements. We plotted the high frequency AC photo-responsivity $R_{AC}$ using the blue dots in the inset of Fig. 2b, and an almost identical curve as the DC one was observed, further supporting the high bandwidth nature of the photo-response.

The mechanism of generation and collection of the photo-carriers is schematically shown in Figs. 3a and b. When light is incident in the high E-field region, electron-hole pairs are generated and separated. In the open-circuit case, these photo-generated carriers are separated by internal fields, but do not produce a current in external circuits. Instead, an open-circuit voltage, $V_{OC}$, is produced across the carrier generation region as shown in Fig. 3a. In the short-circuit case, a photocurrent, $I_{ph} = \dfrac{V_{OC}}{R_g}$, is generated in the external circuit, where $R_g$ is the graphene resistance as shown by Fig. 3b. Unlike in conventional Schottky barrier metal-semiconductor-metal (MSM) photodetectors in which a high bias larger than in the flat-band condition is needed for high bandwidth operation [29], the gap-less nature in single or few-layer graphene allows for the almost unimpeded transmission of carriers through the potential barriers shown in the right side of Fig. 3b, leading to high bandwidth photo-detection even at zero source-drain bias.

To generate a photocurrent in an external circuit, the photo-generated carriers must exit the photo-generation region before they recombine. If we assume that the carrier under strong E-field can travel at a saturation velocity of $5.5 \times 10^5$ m/s at low carrier density [30], it would only take about 0.36 ps for the carriers to travel out of the ~ 200 nm current generation region. Hence, the photo-generated carriers have a good chance to exit the photo-generation region, resulting in reasonably good internal efficiency (6 to 16%) within the high E-field photo-detection region although the lifetime is short (tens of pico-seconds) compared with those in conventional semiconductors. While this short photo-carrier lifetime in graphene is unfavorable for efficient photocurrent generation, it is desirable for other photonic applications such as optical switches, in which quick annihilation of photo-generated carriers is essential for high speed operation.

The transit time limited bandwidth of the photodetector is [20]:

$$f_t = \frac{3.5}{2\pi t_{tr}} \quad (3)$$

where $t_{tr}$ is the transit time through the photo-detection region. Thus, a $f_t$ of 1.5 THz is obtained. If a 200 nm thick $In_{0.53}Ga_{0.47}As$ is used as the photo-detection layer, the transit time limited bandwidth is around 150 GHz, which is about 10 times smaller than that in graphene [20] and mainly limited by the saturation velocity of the holes.

According to the I-V curve shown in Fig. 2a, the graphene photodetector and the high frequency testing circuits can be simulated using the model shown in Fig. 3c. Compared with a conventional photodetector model in which the photodetector is reversely biased and modeled as an ideal current source [20-21], here the graphene photodetector can be modeled using either a current source or a voltage source with an internal resistance $R_g$. In the high frequency $S_{21}$ measurement, the photocurrent signal is transmitted through a 50 Ω transmission line and finally to a matched load $R_L$ of 50 Ω. Here, the total capacitance ($C_p+C_g$) ranges from 27 to 35 fF when the gate bias is in the range of 30 to 80 V. According to this model, the calculated RC limited 3dB bandwidth, $\frac{1}{2\pi R_L(C_g + C_p)}$ is 106 GHz assuming a total capacitance ($C_g+C_p$) of 30 fF. Note that this is not the intrinsic limit of the current graphene photodetector, but limited mainly by the pad capacitance. Measurements of similar, naked G-S pads in the gate bias range of 30 to 80 V yield a pad capacitance of 20 to 25 fF, implying that the $C_g$ of this area of graphene (about 3 μm$^2$) should be 5 to 10 fF. If we assume a graphene capacitance of 5 fF, the RC limited bandwidth will be $\frac{1}{2\pi R_L C_g}$, which is about 640 GHz. Therefore, such graphene photodetectors will be limited mainly by RC since the transit time limited bandwidth is much larger (1.5 THz).

In summary, we demonstrated ultra-high bandwidth photodetectors using single and few layer graphenes. In these novel photodetectors, the interaction of photons and graphene, the properties of the photo-generated carriers, and the transport of the photo-carriers are fundamentally different from those in conventional group IV and III-V materials. These unique properties of graphene enable very high bandwidth (potentially > 500 GHz) light detection, very wide wavelength detection range, zero dark current operation, good internal quantum efficiency and ease of fabrication.

**METHODS**

**Fabrication of the graphene photodetectors**

The graphene photodetectors used in this study were prepared by "mechanical exfoliation" of graphite and are similar to standard back-gated graphene transistors [4, 16-17]. Graphitic flakes were deposited on the surface of a silicon substrate covered by 300 nm thick silicon dioxide and pre-patterned alignment marks. To reduce the parasitic capacitance, a highly resistive wafer (1 to 10 kΩ•cm) was used. Single and few-layer graphenes were identified by visual inspection with an optical microscope and their number was further confirmed by Raman spectroscopy [25]. Electron beam lithography was used to pattern the signal and ground contact electrodes and metal films (Ti/Pd/Au, 0.5/20/20 nm) were deposited using electron-beam evaporation. The electrical contact pads were designed as 50 Ω coplanar transmission lines for on-wafer high-speed characterization. They are 80 × 80 μm wide and are separated by 40 μm. Devices with two types of wirings were fabricated: ground-signal (G-S) and ground-signal-ground (G-S-G) [see Fig. 1a]. The graphene channels are 1 or 2 to 3 layers thick, 1-2 μm long and 1-2.5 μm wide. In the G-S-G configuration, the signal electrode had a width of ~150 nm as shown in Fig. 1a. Results reported in this study are obtained from devices with G-S wirings. Overnight annealing of the as-prepared samples in an ultrahigh-vacuum chamber at 400 K was performed before the measurement in ambient condition. When sweeping the back gate bias, we typically observe the maximum device resistance ("Dirac point") at 20 to 40 V.

**High frequency electrical characterization of the photodetectors**

The electrical characteristics of the photodetectors were inferred from on-chip radio-frequency measurements up to 20 GHz using an Agilent E8364C parameter network analyzer. A standard Short-Open-Load (SOL) calibration substrate (from GGB Industries) was used for calibrating out the losses in the network analyzer, the 3.5-mm cabling, and the microwave probe (GGB Industries) to ensure accurate on-wafer testing. The complex-valued, high-frequency device impedance $Z_{IN}$ was determined by measuring the frequency dependent reflection coefficient $\Gamma_{IN}$ ($S_{11}$ scattering parameter) and calculating $Z$ according to $Z_{IN} = Z_0 \dfrac{1+\Gamma_{IN}}{1-\Gamma_{IN}}$, where $Z_0$ is the 50 Ω system impedance. The lumped circuit model of the photodetector was extracted based on the measurement and subsequent fitting of the model parameters. The gate bias (up to 80 V) was applied using a HP 4145C semiconductor parameter analyzer.

**High frequency optical characterization of the photodetectors**

Optical photodetector characterization was performed using two different sets of Agilent Lightwave Component Analyzers (LCA), N4375B and N4373C, both with a build-in 1550 nm laser. The highest light intensity modulation frequencies available in N4375B and N4373C are 26 GHz and 67 GHz, respectively. Other microwave components in our setup including cables and probes are limited to 40 GHz. The relative photo-response in Fig. 2b was measured using N4373C since it provides a higher modulation frequency. The absolute photo-responses at various gate biases shown in Fig. 2c were measured using N4375B since it has lower noise floor and therefore allows for the measurement of small photo-response signals. Frequency response characterization was achieved by sweeping the light modulation frequency. The optical fiber output of the LCA (-2 dBm) was amplified using an erbium-doped fiber amplifier (EDFA). The EDFA output was then coupled into free space and focused to a diffraction limited spot (~2 μm FWHM diameter) on the device using a 20× magnification, long-working-distance microscope objective. A piezo-electrically driven mirror, mounted before the objective, allowed us to position the beam on the sample with nanometer spatial precision. The total incident optical power on the sample was 3 mW (4.8 dBm). The photocurrent signal was extracted through a G-S microwave probe (GGB Industries) and was fed back into the input port of the parameter network analyzer. The frequency dependent $S_{21}$ scattering parameter, which is defined as $S_{21} = 20 \times \log_{10}\left[\frac{R_{AC}(Amp/Watt)}{1(Amp/Watt)}\right]$ (dB) was recorded as the modulation frequency was swept between 10 MHz and 26 GHz. From this measurement, the high frequency AC responsivity $R_{AC} = \Delta I_{ph}/\Delta P_{in}$ was then determined. The impact of the EDFA on the responsivity is calibrated out in the final data. The gate bias was applied using a HP 4145C semiconductor parameter analyzer.

**Acknowledgments**

The authors are grateful to M. Freitag and Z. Chen for helpful discussions, to Y. Vlasov, S. Assefa, W. Green, C. Schow and L. Schares for help with the RF measurements, to J. Tsang for Raman measurements, and to B. Ek and J. Bucchignano for help in technical assistance. F. X. is indebted to C. Y. Sung for his encouragement. T. M. acknowledges financial support by the Austrian Science Fund (FWF).

**Figure Captions**

**Figure 1 Electrical characterizations of the graphene photodetector**

a) SEM and optical (inset) images of the high-bandwidth graphene photodetectors. The graphene shown here has 2 to 3 layers. Two types of wirings are shown: Ground-Signal (G-S) and Ground-Signal-Ground (G-S-G). The high frequency results are from devices with G-S wirings. Scale bar (inset): 2 μm (80 μm).

b) Device schematics and electrical model in the high frequency domain. High frequency reflection coefficient $\Gamma_{IN}$ is measured using a network analyzer, and $Z_{IN}$ is inferred from $\Gamma_{IN}$. The green symbols from top to bottom represent $C_p$, $C_g$, and $R_g$, respectively. The purple sheet denotes the graphene, and a pair of dark red strips denotes the microwave probe tips.

c) Measure circuit parameters in DC and high frequency domains at different gate biases. Solid blue: DC resistance. Blue and red dots: resistance and total capacitance in high frequency domain from reflection measurement.

**Figure 2 Optical characterizations of the graphene photodetector**

a) Typical I-V curves of the graphene photodetector without and with light excitation. Inset: schematic of the photocurrent measurement. The curved arrow in the inset represents the incident photon.

b) Relative AC photo-response $S_{21}(f)$ as a function of light intensity modulation frequency up to 40 GHz at a gate bias of 80 V. $S_{21} = 20 \times \log_{10}\left[\dfrac{R_{AC}(Amp/Watt)}{1(Amp/Watt)}\right]$ (dB), where $R_{AC} = \Delta I_{ph}/\Delta P_{in}$ (A/W). The curve is displaced vertically hence the response starts from 0 dB. Response degradation of ~1 dB is observed at 40 GHz, which is not due to the graphene photo-response but the microwave probes. Inset: Peak DC and high frequency (AC) photo-responsivity as a function of gate bias.

c) Absolute AC photo-response $S_{21}(f)$ as a function of light intensity modulation frequency up to 26 GHz at a gate bias of 80 V. Inset: Absolute AC photo-response $S_{21}$ (f) measured at a number of gate biases varying from -40 to 80 V. Here, $S_{21}$ and $R_{AC}$ are defined the same as in Fig. 2b. However, absolute values are presented here. Response degradation of ~0.5 dB at 26 GHz is not caused by the graphene device as in Fig. 2b. AC photo-responsivity obtained here is presented in the inset of Fig. 2b.

**Fig. 3 Graphene photodetector circuit model and analysis**

a) Potential profile and open-circuit voltage generation mechanism in a graphene photodetector operated under open-circuit condition.

b) Potential profile and photocurrent generation mechanism in a graphene photodetector operated under short-circuit condition. Since photo-generation region is much smaller than the total graphene channel length (0.2 vs. 1.2 µm), the resistance within the photo-generation region is ignored and channel resistance $R_g$ is used to represent the resistance of the rest of the channel. In both (a) and (b), the curved red arrow denotes the incidence photon, the black solid and hollow dots represent the photo-generated electron and hole, respectively, the solid grey line denotes the potential within the graphene channel, the dashed blue line shows the Fermi level, the violet rectangular boxes denote the metal contact, and the blue and red parabolic curves represent a bi-layer graphene band structure.

c) High-frequency equivalent circuit model of the graphene photodetector. Components to the left of the dashed line ($i_{ph}$, $R_g$ and $C_g$) represent the graphene device, while those to the right denote pad capacitance ($C_p$), 50 Ω transmission line, and 50 Ω load. In the DC photocurrent measurement, $R_L = 0$ Ω and short-circuit photocurrent is measured. In high frequency case, $R_L$ (50 Ω) is still much smaller than the internal resistance $R_g$, therefore the detector still operates at close to short-circuit condition.

Figure 1

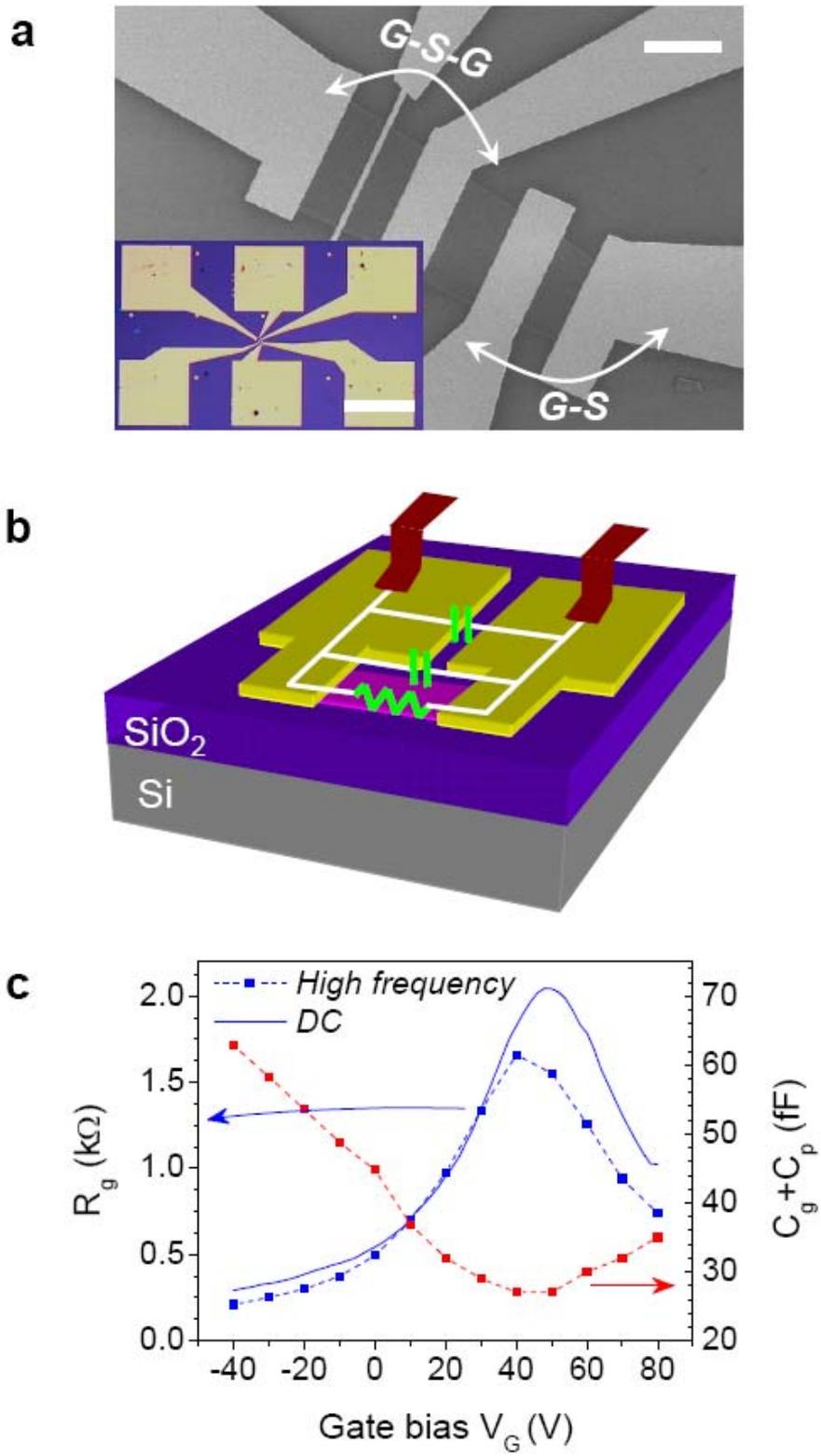

Figure 2

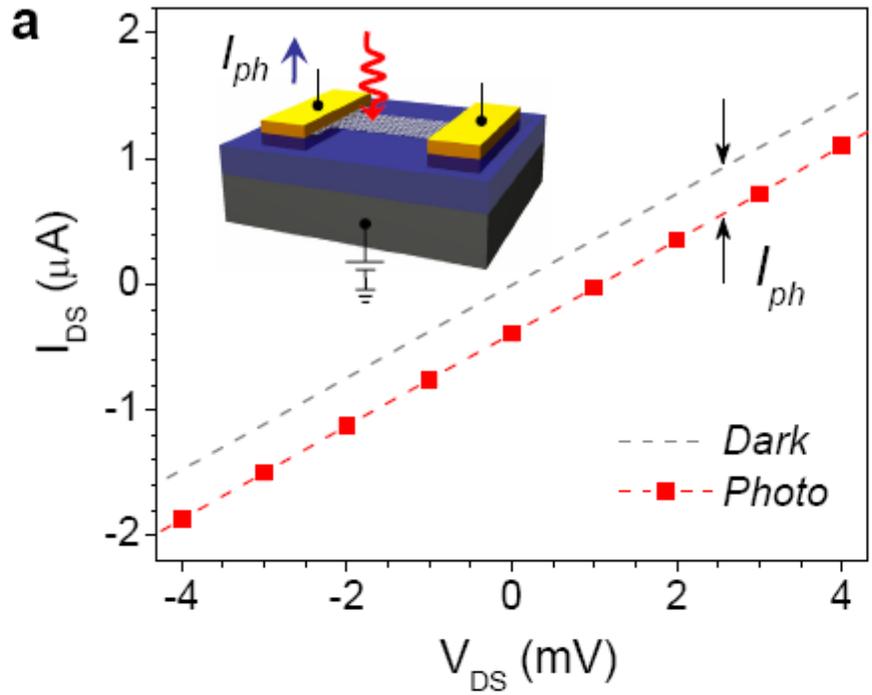

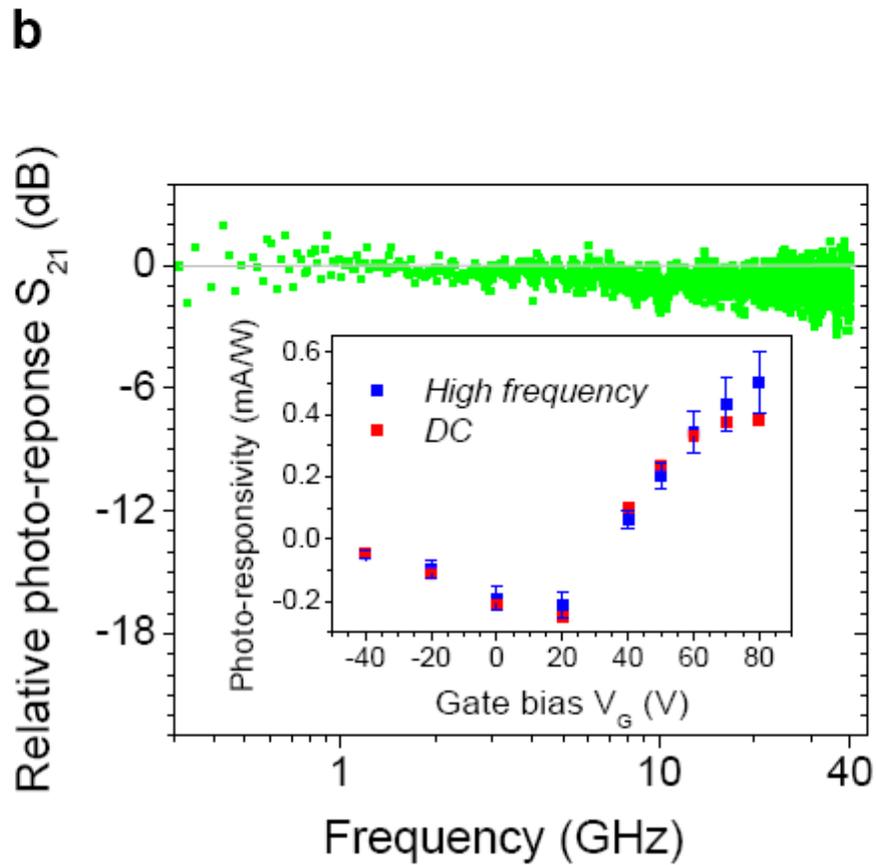

Figure 2

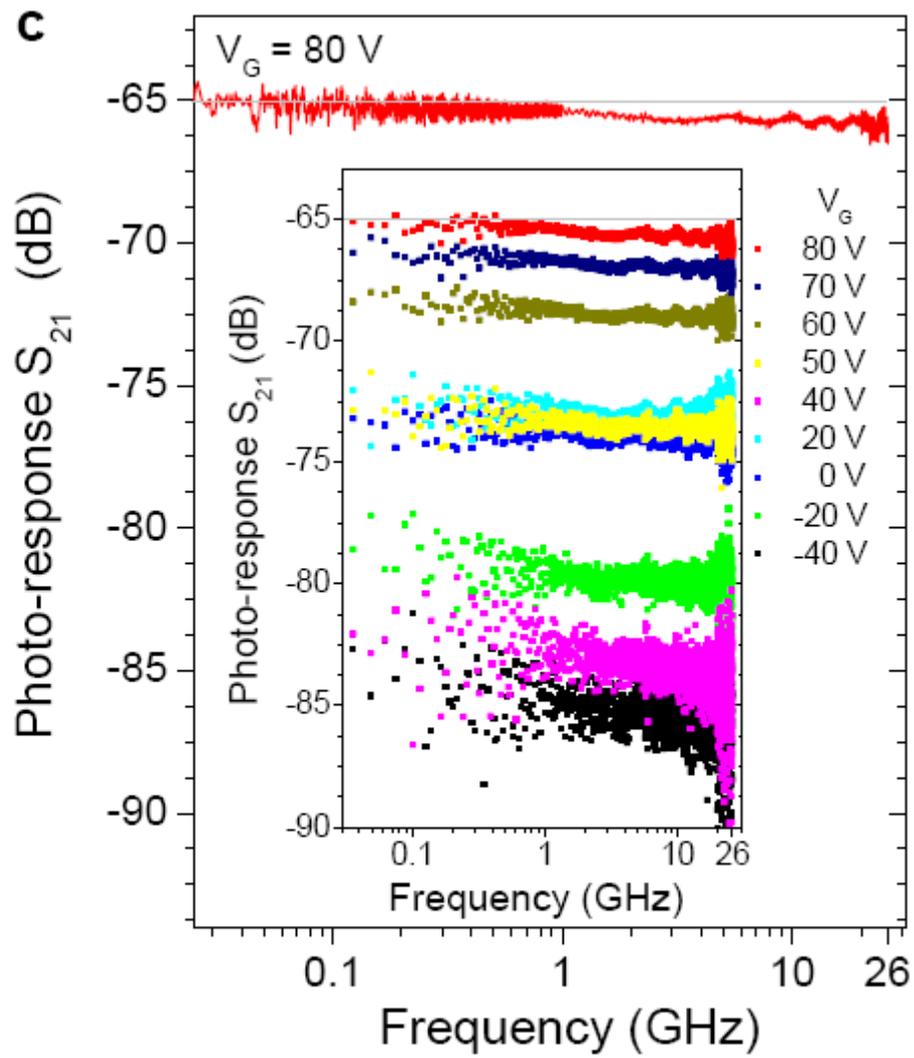

Figure 3

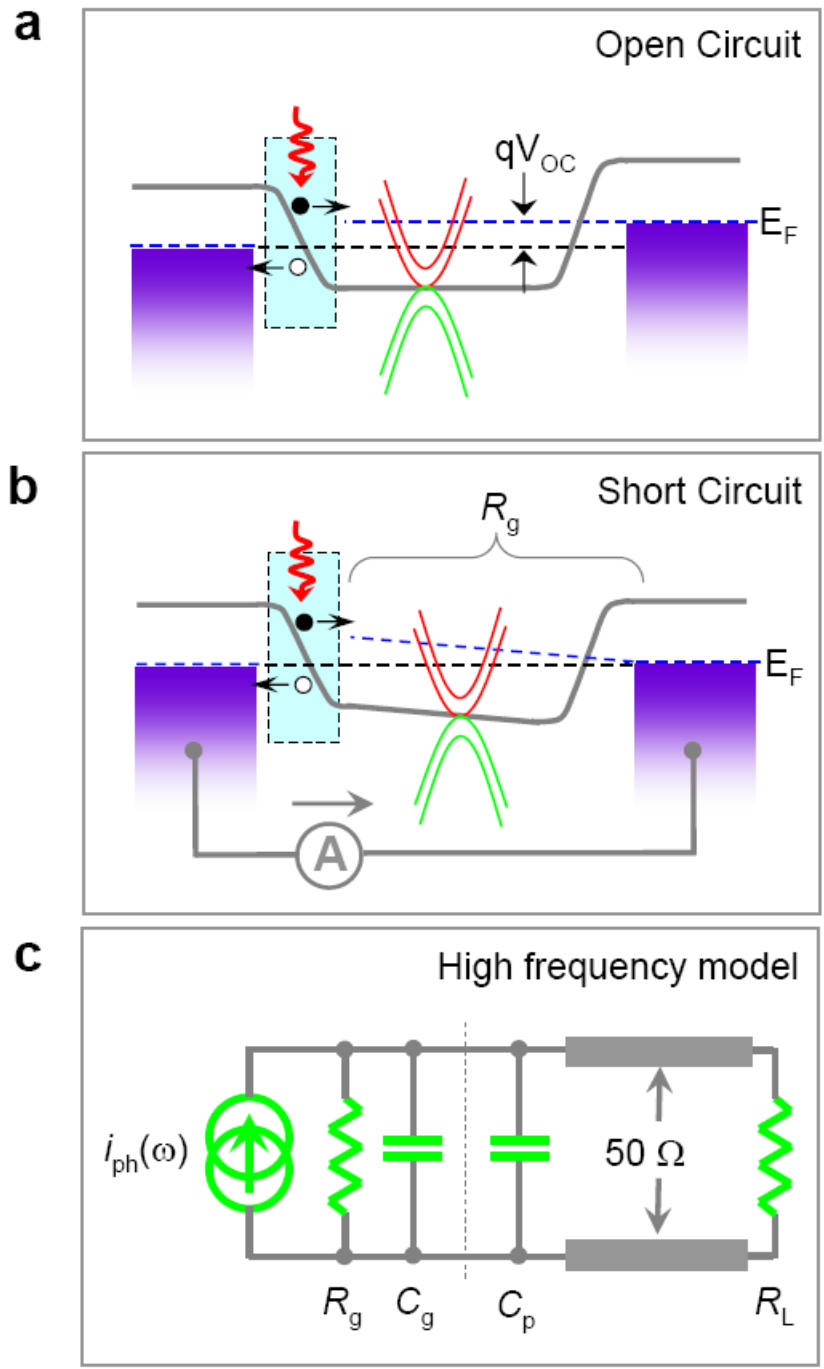

# Supplementary Information

# Discussion on the internal and external quantum efficiency of graphene photodetectors

Fengnian Xia[*‡], Thomas Mueller[‡], Yu-ming Lin, Alberto Valdes-Garcia and Phaedon Avouris[*]

www.research.ibm.com/nanoscience

*IBM Thomas J. Watson Research Centre, Yorktown Heights, NY 10598*

**I. Internal and external quantum efficiency in few-layer graphene photodetectors**

The reported external efficiency of our device is not intrinsic to the graphene photodetector and can be improved considerably. There are two major factors which limit the external efficiency. The first one is the limited absorption of the light within few-layers of graphene in the current device configuration. A suspended bi- or tri-layer graphene in air absorbs about 4.6% and 6.9% of the vertical incidence light, respectively. In this experiment, the graphene is on top of a 300 nm thick oxide on silicon substrate. At 1.55 μm, although the absorption is estimated to be enhanced by about 25%, it is still far from complete absorption. The second is the width of the effective photo-detection area of the graphene device (or high built-in E-field region), which is only about 100 to 250 nm long (determined by the charge transfer between the metal and graphene) [S1-S2]. Given that the light spot size is around 2 μm in diameter, the estimated internal quantum efficiency (the ratio of the number of photo-carriers collected in the external circuits to the number of photons actually absorbed in the small graphene photo-detection area) within the high E-field photo-detection region is 6 to 16%. The efficiency can be enhanced by creating a wider photo-detection region [S1-S2] using split gates, a longer light-graphene interaction length through waveguide [S3] or cavity integration, or simply by reducing the internal resistance $R_g$. Furthermore, if a bi-layer graphene is integrated on top of a 550 nm by 220 nm silicon optical waveguide, our numerical simulation indicates that the TM (transverse magnetic) light in such an integrated system can be absorbed by 65% after propagating in the waveguide for 20 μm. In recent experiments on graphene p-i-n junctions, we also found considerable improvement in external responsivity (efficiency). Detailed discussions about the schemes and results of improving the external responsivity of graphene photodetectors will be presented elsewhere.

**II. DC and high frequency photo-response in single-layer graphene**

The photocurrent measurements reported in the main text are obtained from few-layer graphene devices (bi- or tri- layer). In addition, we also performed high frequency photocurrent measurements on single-layer graphene field-effect-transistor (FET) based photodetectors. The results are presented in Fig. S1.

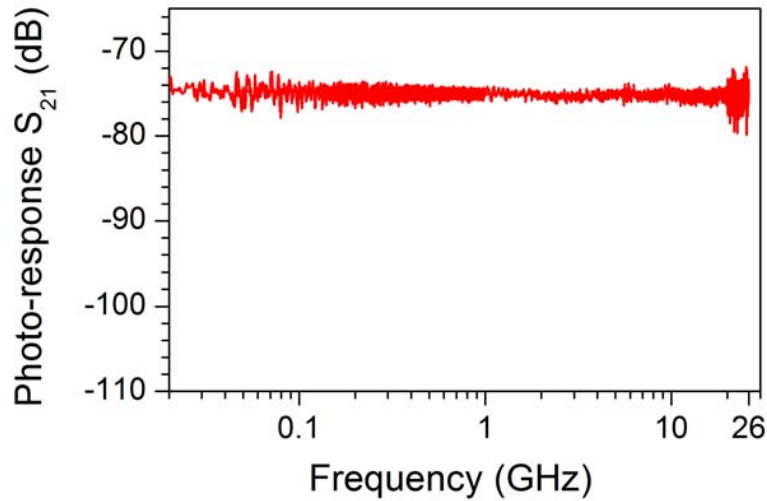

Fig. S1. Peak high frequency photo-response $S_{21}$ as a function of light intensity modulation frequency at a gate bias of 20 V in a single layer graphene device.

No photoresponse degradation is observed up to 26 GHz light intensity modulation frequency. As presented in the main text, here $S_{21} = 20 \times \log_{10} \frac{R_{AC}(Amp/Watt)}{1(Amp/Watt)}$, where $R_{AC} = \frac{\Delta I_{ph}}{\Delta P_{in}}$ (A/W). From Fig. S1, we calculate a maximum AC external responsivity value of 0.2 mA/W at 20 V gate bias, about 40% of the maximum value reported in the main text in which a few-layer graphene device is reported. Similar maximum DC external responsivity of around 0.2 mA/W is observed in this single layer graphene device. In this device, the maximum responsivity is achieved at different gate bias if compared with the device presented in the main text due to the variation of Dirac point voltage.

‡*These authors contributed equally to this work*
**Email:* fxia@us.ibm.com; avouris@us.ibm.com*